\documentclass[onecolumn,runningheads,amssymb,amsmath,natbib]{svjour2}
\usepackage{epsfig}
\usepackage{graphicx}
\usepackage{epic}
\usepackage{eepic}
\usepackage{bbold}
\usepackage{mathptmx}
\journalname{Celestial Mechanics and Dynamical Astronomy}

\newcommand{\be}{\begin{equation}}
\newcommand{\ee}{\end{equation}}
\newcommand{\ba}{\begin{eqnarray}}
\newcommand{\ea}{\end{eqnarray}}

\begin{document}

\title{Phase-Space Volume of Regions of Trapped Motion: Multiple Ring
  Components and Arcs}%
\titlerunning{Phase-Space Volume of Regions of Trapped Motion}%
\author{Luis Benet \and Olivier Merlo}%
\institute{Luis Benet \and Olivier Merlo%
    \at%
    Instituto de Ciencias F\'{\i}sicas, Universidad Nacional Aut\'onoma
    de M\'exico (UNAM)\\
    Apdo. Postal 48--3, 62251--Cuernavaca, Mor., M\'exico\\
    \email{benet@fis.unam.mx, merlo@fis.unam.mx}%
  }%

\date{\today}

\maketitle

\begin{abstract}
  The phase--space volume of regions of regular or trapped motion, for
  bounded or scattering systems with two degrees of freedom
  respectively, displays universal properties. In particular, sudden
  reductions in the phase-space volume or gaps are observed at
  specific values of the parameter which tunes the dynamics; these
  locations are approximated by the stability resonances. The latter
  are defined by a resonant condition on the stability exponents of a
  central linearly stable periodic orbit. We show that, for more than
  two degrees of freedom, these resonances can be excited opening up
  gaps, which effectively separate and reduce the regions of trapped
  motion in phase space. Using the scattering approach to narrow rings
  and a billiard system as example, we demonstrate that this mechanism
  yields rings with two or more components. Arcs are also obtained,
  specifically when an additional (mean-motion) resonance condition is
  met. We obtain a complete representation of the phase-space volume
  occupied by the regions of trapped motion.
\end{abstract}

\section{Aims, motivation and scope}

The aim of this paper is two fold: On the one hand, we address the
question of the dependence of the phase-space volume of the regions of
trapped motion in terms of a parameter that tunes the dynamics. The
class of systems we address are dominated by escaping trajectories,
i.e. scattering systems, which may exhibit dynamically trapped
motion. Examples include H\'enon's map, open billiard systems,
scattering maps, the RTBP, among others, i.e., Hamiltonian scattering
systems with two and more degrees of freedom. This question was first
formulated and answered for bounded Hamiltonian systems of two degrees
of freedom in terms of the regions in phase space that exhibit ordered
motion~\citep{Contopoulos2005,Dvorak2005}. For two degrees of freedom
Hamiltonian systems, the volume in phase-space of the regions of
trapped or ordered motion displays universal behavior in terms of a
parameter that controls the
dynamics~\citep{Contopoulos2005,Dvorak2005}, revealing abrupt changes
at specific locations which can be well approximated~\citep{Benet2008}
and exhibiting self-similar
behavior~\citep{SimoVieiro2007}. Universality means that the same
behavior and predictions are obtained for a wide class of
systems. Yet, for systems with more than two degrees of freedom there
are no results of this type, which therefore calls for attention. The
evident difficulty lies in the higher dimensionality of phase space,
which prevents the use of standard techniques to distinguish whether
an initial condition belongs to a phase space region of ordered or
trapped motion or not. We therefore shall study the dependence upon a
parameter that tunes the dynamics of the phase space volume of the
regions of trapped motion in a specific scattering system with two or
two-and-half degrees of freedom.

The second aspect that we address here are the implications of the
previous study in the formation of dynamical patterns obtained by
projecting the phase-space locations of many non-interacting particles
at a given time into the configuration space (real space); we refer to
these patterns as rings. More specifically, we consider the appearance
of fine structure and its azimuthal dependence in the context of the
narrow planetary rings, which are formed by the ensemble of
non-escaping particles. Our study is based on the scattering
approach~\citep{Merlo2007}, and we use as illustrative example an
unrealistic system, namely, a planar circular hard disk whose center
moves along a Keplerian closed trajectory. Despite of the
non-realistic interactions considered in this example, we do obtain
rings which display more than one component and arcs, features which
are in qualitative agreement with observations of real narrow
planetary rings (for an account of the observations see the recent
book by~\citealt{Esposito2006}). The important aspect is not how
realistic is the example used, but the fact that the occurrence of
structure in these rings, such as multiple components and arcs, is
understood and explained in terms of the underlying phase space. The
additional costs of integrating more complicated equations of motion
may not provide any additional conceptual understanding; this was
actually the main motivation for the introduction of billiard systems
by~\citet{Birkhoff1927}.

The reader may ask what is the relevant contribution of our work for
the understanding of real planetary rings, bearing in mind that we use
a non-realistic system with non-interacting ring particles as example
to illustrate our ideas. Indeed, planetary rings are flat
self-gravitating systems consisting of many colliding 
particles~\citep[cf. chapter 4]{Esposito2006}, which
revolve around oblate planets and are perturbed by nearby
satellites. The connection of the billiard system with the restricted
three-body problem was formally established by~\citet{Benet2001} and
explicitly illustrated in~\citet{Merlo2007}. Briefly, one first
includes the central planet at the origin and its gravitational
attraction, and then considers the limit of vanishing radius of the
disk. It can be shown that this situation is identical to the restricted
three-body problem with mass parameter $\mu=0$~\citep{Henon1968}; in
particular, we note that both problems share the simple
consecutive-collisions periodic orbits. \citet{Hitzl1977} proved that
the second species periodic orbits of the restricted three-body
problem when $\mu\to 0$ are precisely the critical points of the
consecutive-collision periodic orbits for $\mu=0$, which correspond to
the vanishingly small disk with a central $1/r$ attractive
potential. This shows that our results using the rotating billiard
system are indeed connected with more realistic models such as the
restricted three-body problem, and that the underlying phase-space
structure of these problems is similar. While this point is certainly
important, more relevant is the fact that this shows the robustness of
the scattering approach. Robustness warranties that we can extend the
model beyond the restricted three-body problem, and include the
oblateness of the planet as well as other perturbations, e.g., the
gravitational influence of other satellites. We refer the reader
to~\citet{Merlo2007}, where these and other physically relevant
aspects of the scattering approach, in particular the predictions in
connection to some structural properties of narrow rings, are
discussed in detail.

The paper is organized as follows: In the next section we briefly
describe the toy-model (or example) we use throughout the paper, a
scattering billiard system that moves on a Kepler circular or elliptic
orbit. In section~\ref{sec:rings} we obtain the rings and their structural
properties in terms of the eccentricity of the orbit of the center of
the disk. In particular, we obtain rings with two or more components
which in addition may reveal arcs.  In section~\ref{sec:ps_volume} we
relate the fine structure observed in the rings with the specific
behavior of the phase space volume of the regions of trapped motion
in terms of a parameter that controls the dynamics. Finally, in
section~\ref{sec:concl} we summarize and provide our conclusions.

\section{The scattering billiard on a Kepler orbit}
\label{sec:toy-system}

In this section we briefly describe our toy model, a disk on a Kepler
orbit, which is a scattering billiard system~\citep{Meyer1995}. This
is a simple yet compelling toy model for chaotic
scattering~\citep{Dullin1998}, which includes the case of more than
two degrees of freedom~\citep{Benet2005}. In addition, it is the
simplest example that illustrates the occurrence of narrow rings
within the scattering approach and allows some analytical
treatment~\citep{Benet2004,Merlo2007}.

The system consists of a point particle (ring particle) moving freely
on the $X$-$Y$ plane which may collide with an impenetrable circular
hard-disk of radius $d$, whose center moves on a Kepler
ellipse. Bounces with the disk are treated as usual in the local
(moving) reference frame at the point of collision~\citep{Meyer1995};
particles that do not bounce with the disk escape inevitably to
infinity along scattering trajectories. In the context of rings we are
interested in the initial conditions of particles that stay trapped in
the system for very long if not infinitely long times. Therefore,
bounces with the disk are the only way in which the potential may
confine the motion of the ring particles. Note that bounces are
necessary, but not sufficient, to avoid escaping to infinity. Hence,
we shall focus on the phase space conditions that ensure trapping for
a set of initial conditions of positive measure.

\begin{figure}
  \centering \includegraphics[angle=90,width=6cm]{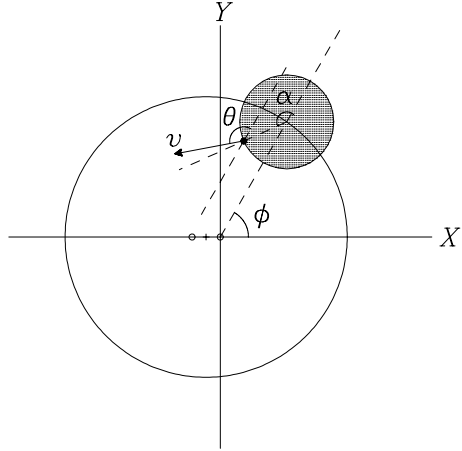}%
  \caption{ The geometry of the scattering billiard on a Kepler orbit:
    $\phi$ is the angular position of the center of the disk, $\alpha$
    denotes the position of the collision point on the disk, $v$ is
    the magnitude of the outgoing velocity and $\theta$ defines its
    direction. The center of the disk moves on a Kepler ellipse, whose
    foci are shown on the $X$-axis as open circles ($\circ$), and its
    center by a cross~($+$). The shaded circular region is the disk.}
  \label{fig:geom}
\end{figure}

Figure~\ref{fig:geom} illustrates the geometry of the scattering
billiard on a Kepler orbit. The center of the disk moves on a circular
or elliptic Kepler orbit of semi-major axis $R=1$ ($R>d$), the origin
being located at one of the foci of the ellipse. We denote by
$\vec{X}_d(\phi)$ the position in an inertial frame of the center of
the disk in terms of $\phi$, the true anomaly (measured from
pericenter), $R_d(\phi)$ denotes the radial component and
$\varepsilon$ is the eccentricity of the orbit. The Hamiltonian of
this system in an inertial frame is given by
\begin{equation}
  \label{eq:ham}
  H=\frac{\vec{P}^2}{2}+V_d(|\vec{X}-\vec{X}_d(\phi(t))|)\ .
\end{equation}
Here, $\vec{X}$ and $\vec{P}$ are the coordinates and the canonically
conjugated momenta of the ring particle, respectively, and
$V_d(|\vec{X}-\vec{X}_d(\phi(t))|)$ is the interaction potential with
the disk. The latter is zero if $|\vec{X} - \vec{X}_d(\phi(t))| > d$
and infinite otherwise.

For non-zero $\varepsilon$, due to the explicit time dependence, the
Hamiltonian system has two-and-half degrees of freedom and no constant
of motion; for $\varepsilon=0$ there is a constant of motion, the
Jacobi integral, and the system has only two degrees of freedom. The
Jacobi integral is the Hamiltonian expressed in a rotating frame. In
general, we may parameterize the phase space of the system by the
usual coordinates and momenta for the particle and $\phi(t)$ for the
position of the disk. Yet, since we are interested in the case where
the particle is dynamically trapped through collisions with the disk,
a simpler parameterization is given as follows. Let $\alpha$ denote
the angular position of the collision point on the disk, $v$ the
magnitude of the outgoing velocity after the collision, and $\theta$
be the angle defining the outgoing direction of the velocity with
respect to the vector $\vec{X}_d(\phi(t))$. We denote by $t_i$ the
time when the $i$-th collision takes place; whenever we shall need to
refer to a specific initial condition, we shall use the notation
$t_i^{(k)}$. Notice that the precise outcome of a collision with the
disk depends upon the position on the disk where it occurs, the
relative velocities, and for non-vanishing $\varepsilon$, on the
position on the ellipse where the disk is located at the collision
time, $\phi(t)$. For circular motion of the disk, collisions taking
place on the front of the disk increase the (outgoing) kinetic energy
of the particle, while collisions on the back reduce it. This allows
to construct explicitly the simple periodic orbits.

\section{Narrow rings and fine structure}
\label{sec:rings}

We are interested in the structural properties of a ring, whenever it
occurs. Its occurrence can be understood in simple and general terms
within the scattering approach to narrow
rings~\citep{Merlo2007}. Basically, we consider the scattering
dynamics of an ensemble of non-interacting ring particles that move
under the influence of an intrinsic rotation. For the planetary case,
the intrinsic rotation corresponds to the time-dependent forces
(assumed periodic or quasi-periodic) that originate from the motion of
the planetary moons that orbit around the central planet; a solution
of the $N$-body problem for the planet and moons defines a restricted
($N+1$)-body problem to describe the motion of the ring-particle. The
intrinsic rotation creates generically regions in the extended phase
space of dynamically trapped motion close to the linearly stable
periodic orbits, in what otherwise is dominated by escaping
trajectories~\citep{Benet2000,Benet2001}. These regions are in general
quite small and localized, appearing only in certain bounded regions
of the extended phase space. We set the initial conditions of the
ensemble of non-interacting particles, the ring particles, in a domain
that contains these regions of trapped motion, and let the system
evolve for a time which is long enough. Particles that do not belong
to these regions escape in short time scales. For the particles that
remain trapped, we project their phase-space locations onto the $X-Y$
plane. The pattern formed is a ring. The ring is typically narrow,
sharp--edged and non-circular. These properties follow from the
phase-space regions of trapped motion which are rather localized in
phase space, the scattering dynamics, and the shape of the organizing
centers (periodic orbits) of the regions of bounded motion,
respectively~\citep[see][]{Merlo2007}.

The billiard on a Kepler orbit is a good example for illustrating the
scattering approach: It allows to certain analytical treatment and the
integration of the equations of motion for the billiard system is
essentially exact. For the disk on a circular orbit, $\varepsilon=0$,
the organizing centers of the dynamics can be explicitly calculated
and correspond to the radial consecutive-collision periodic orbits,
i.e., those where the kinetic energy stays constant after the
collision~\citep{Meyer1995}. These radial consecutive-collision orbits
can be described in terms of the Jacobi integral $J$ and the angle
$\theta$ defined above as~\citep{Merlo2007}%
\begin{equation}
  \label{eq:jac}
  J =  2\omega_d^2 (R-d)^2 
  (1+\Delta\phi\tan\theta) \cos^2\theta(\Delta\phi)^{-2}.
\end{equation}
Here, $\Delta\phi=(2n-1)\pi+2\theta$ is the angular displacement of
the disk between consecutive radial collisions and $n=0,1,2,\dots$ is
the number of full rotations completed by the disk before the next
collision. The period of the motion of the disk is $T_d=2\pi$
($\omega_d=1$). We note that the form of Eq.~(\ref{eq:jac}) implies
that the radial collision periodic orbits appear, in terms of $J$,
through saddle-center bifurcations, i.e., in pairs, with one linearly
stable and the other linearly unstable close enough to the bifurcation
point~\citep{Benet2000}; the bifurcation point is defined by the
condition ${\rm d}J/{\rm d}\theta=0$.

For systems with two degrees of freedom, the dynamics can be analyzed
using standard methods, e.g., with a Poincar\'e section. In
particular, for $\varepsilon=0$ we consider the symplectic map of the
surface of section onto itself, $(\alpha_{k+1},p_{k+1})={\cal
  P}_J(\alpha_k,p_k)$, defined when the particle collides with the
disk, where $p_k=-d-R \cos\alpha_k - v_k \sin(\alpha_k-\theta_k)$ is kin
to the angular momentum~\citep{Merlo2007}. The map is constructed by
solving numerically the transcendental equation (using Newton's method
to high accuracy) which determines the time for the next collision
(see~\citealt{Benet2005} for details); whenever there is a solution,
the resulting outgoing variables are computed. Note that this map is
smooth and locally invertible. Linear stability of the radial
collision periodic orbits is explicitly given by the trace of the
linearized dynamics $ D{\cal P}_J$, which can be obtained
analytically~\citep{Benet2004,Merlo2004},
\begin{equation}
  \label{eq:stab}
  {\rm Tr\,} D{\cal P}_J  =  2 +
  \big[(\Delta\phi)^2(1-\tan^2\theta)-4(1+\Delta\phi\tan\theta)\big]R/d.
\end{equation}
Then, the radial collision periodic orbits are {\it linearly} stable
iff $|{\rm Tr\,} D{\cal P}_J|\le2$.

\begin{figure}
  \centering
  \includegraphics[angle=0,width=10cm]{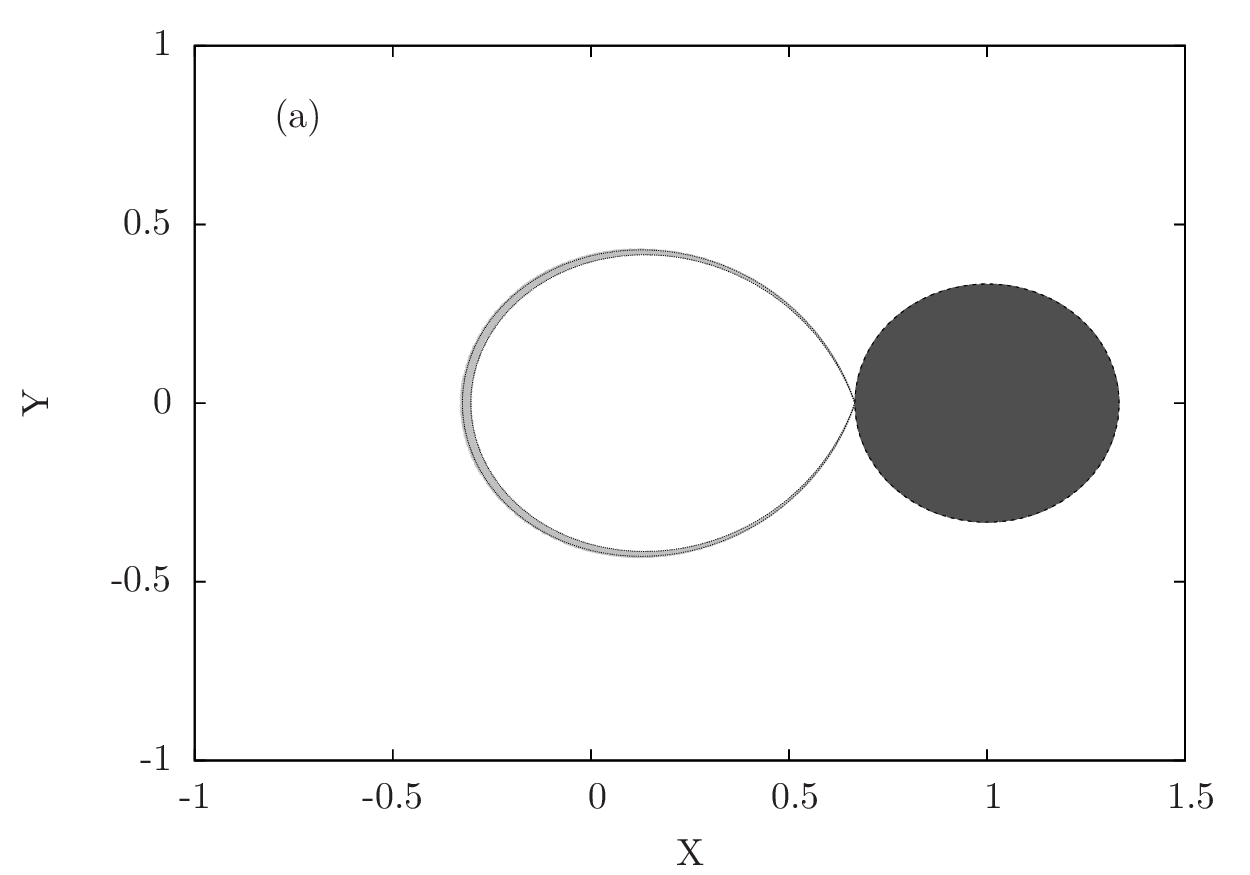} \hspace*{-20pt}
  \includegraphics[angle=0,width=10cm]{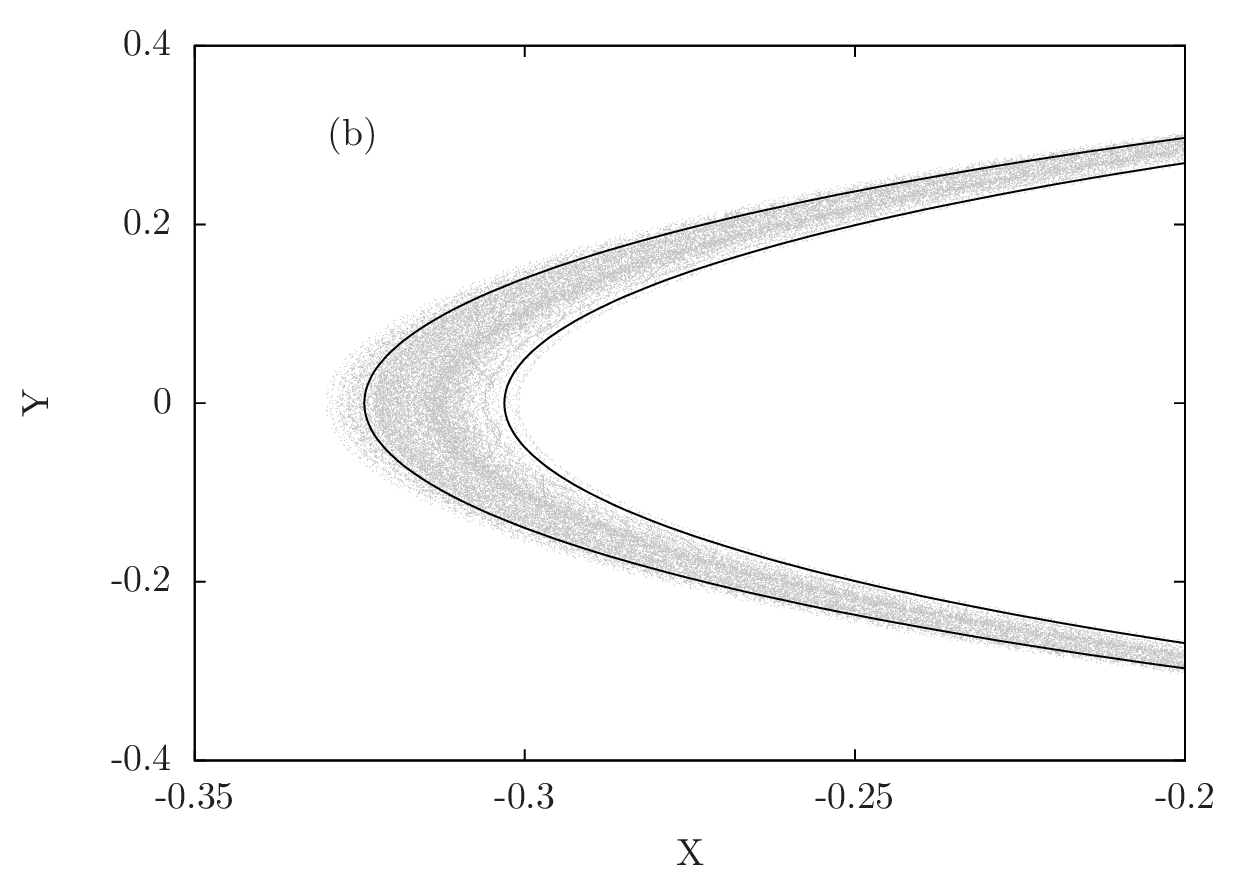}%
  \caption{ (a)~Stable ring of non-interacting particles of the
    billiard on a circular orbit for the $n=0$ stable periodic orbit
    at a given time. The dark shaded circular region is the
    disk. (b)~Detail of a region of the ring. The black lines are the
    analytical estimates given by ${\rm Tr\,} D{\cal P}_J = \pm 2$. }
  \label{fig:ring0} 
\end{figure}

For the case of two degrees of freedom, $\varepsilon=0$, close enough
to the linearly stable radial collision periodic orbits KAM
theorem~\citep{Arnold1988,Nekhoroshev1977} implies that there exist
invariant tori where the motion is quasi-periodic and the set has a
positive measure. Note that KAM theorem applies since the map is a
Hamiltonian diffeomorphism for almost all $J$ and the Diophantine
condition is fulfilled; actually the theorem can be applied for all
values of $J$ where the linearly stable radial periodic orbits has
complex eigenvalues $\lambda$ in the unit circle such that
$\lambda^2\neq 1$, $\lambda^3\neq 1$, $\lambda^4\neq 1$ and the first
coefficient of the resonant normal form is
non-zero~\citep{Duarte1994}. Clearly, the KAM-tori are trapped orbits,
which also holds for the regions of chaotic motion that are bounded by
those KAM tori. More important, the invariant manifolds of the
external unstable fixed point define a bounded region, which contains
the stable fixed point. This region contains a domain where the motion
is strictly bounded or dynamically trapped, i.e., where the
corresponding trajectories cannot escape to infinity. We consider the
whole interval of $J$ where such regions exist; for concreteness we
focus in the $n=0$ radial collision periodic orbit,
cf.~Eq.~(\ref{eq:jac}), which corresponds to the largest region of
trapped motion in phase space. We proceed as described above to obtain
the ring, which is illustrated in figure~\ref{fig:ring0}. In the
figure we included analytical estimates for the borders based on the
projection of the curves ${\rm Tr\,} D{\cal P}_J=\pm 2$, which are the
limits for having linearly stable radial-collision periodic
orbits. The resulting ring is narrow, eccentric, displays sharp-edges
and the estimates based on linear stability are excellent.

\begin{figure}
  \centerline{ 
     \includegraphics[width=8cm]{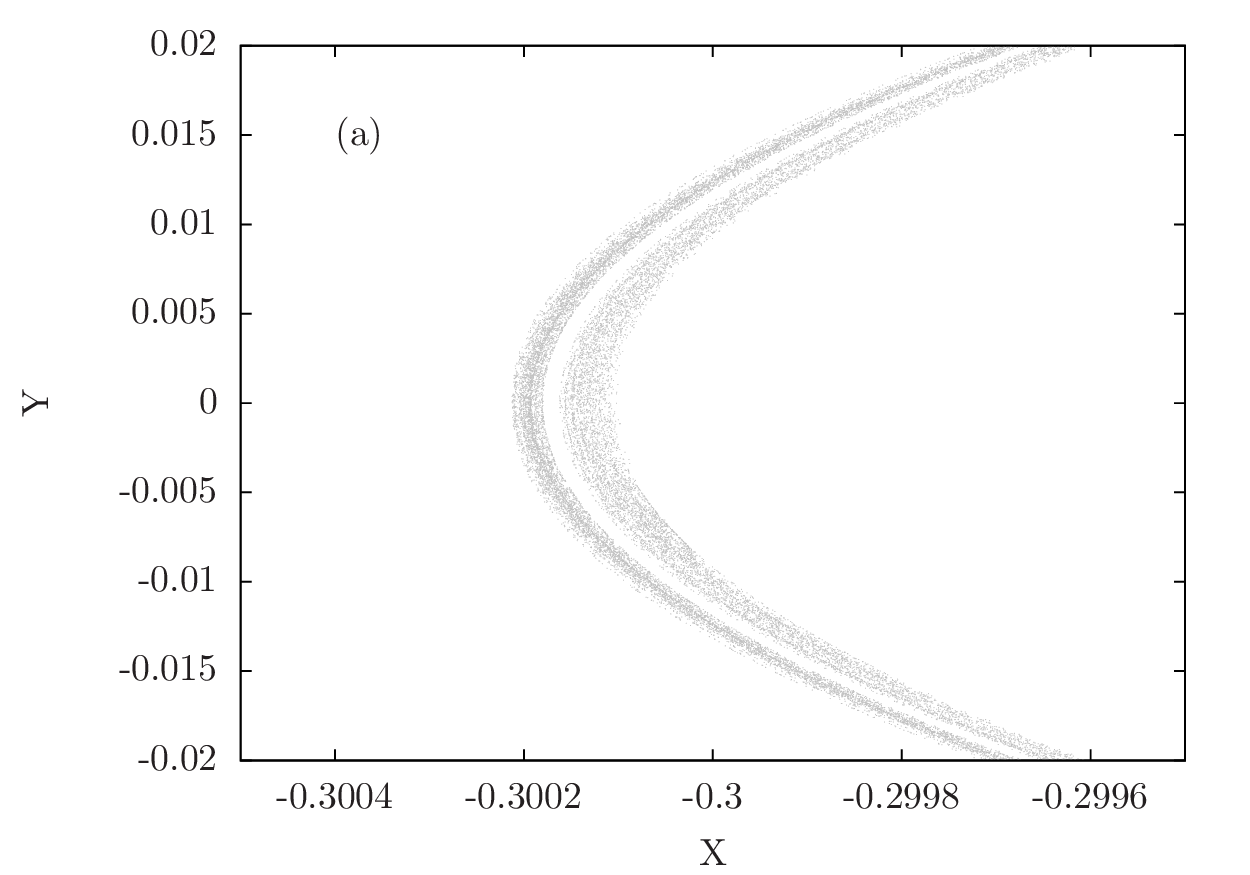}%
     \includegraphics[width=8cm]{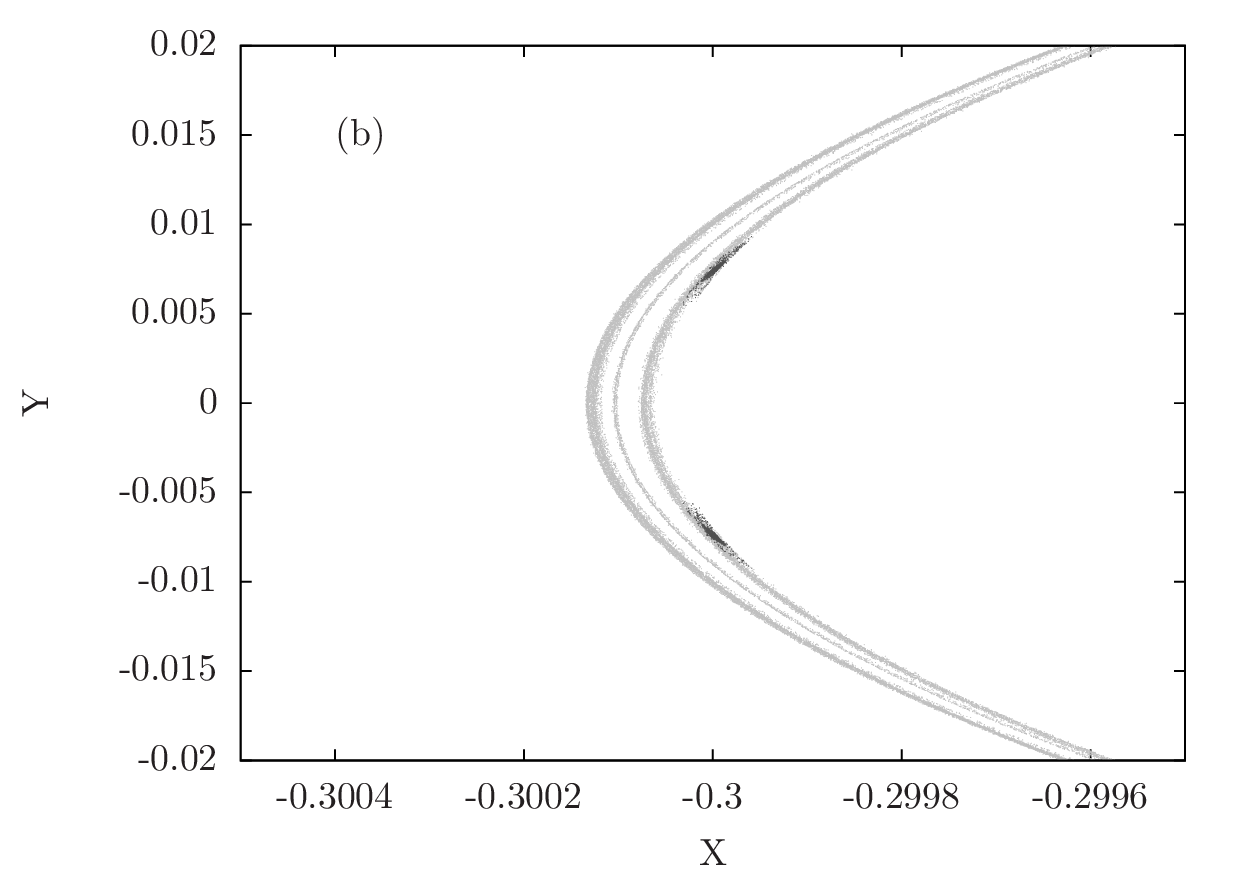}}%
  \centerline{
    \includegraphics[width=8cm]{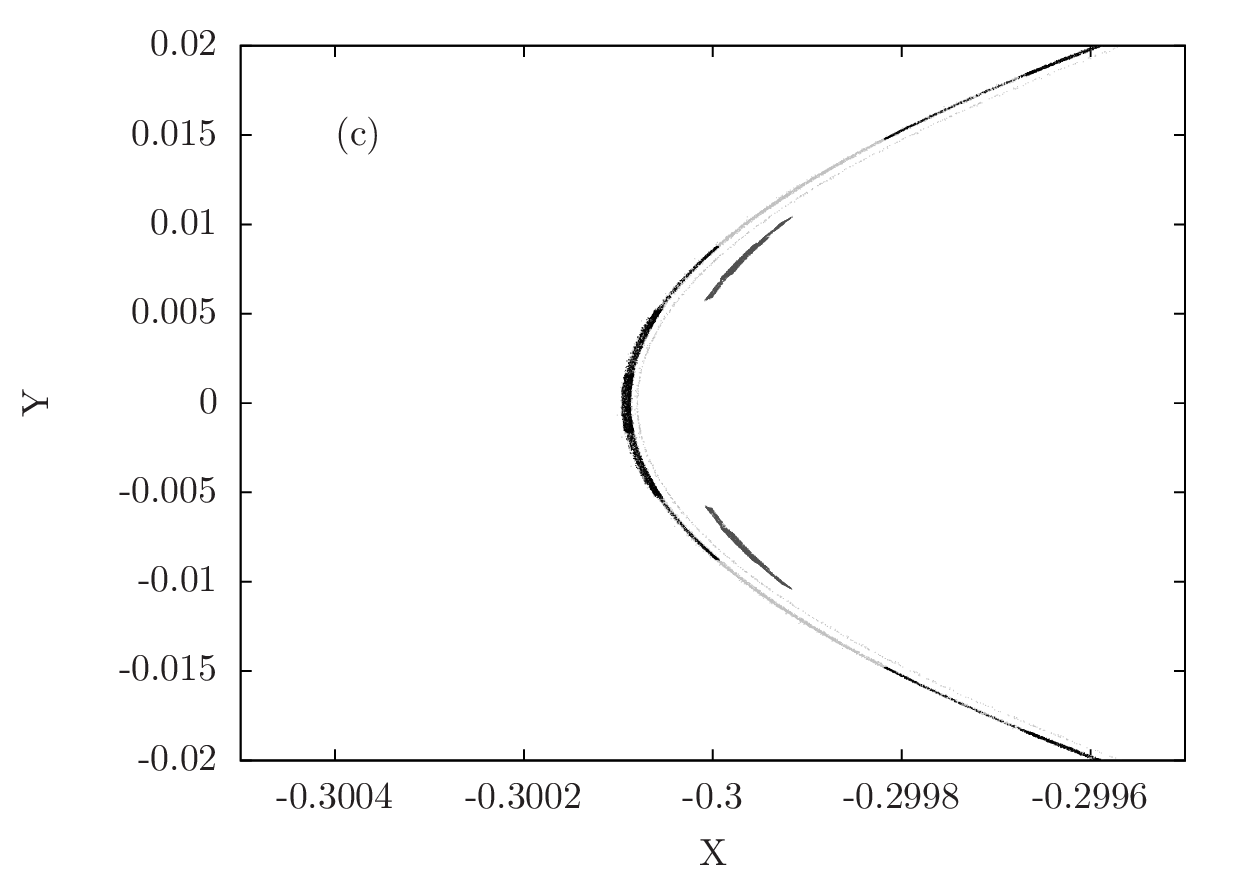}%
    \includegraphics[width=8cm]{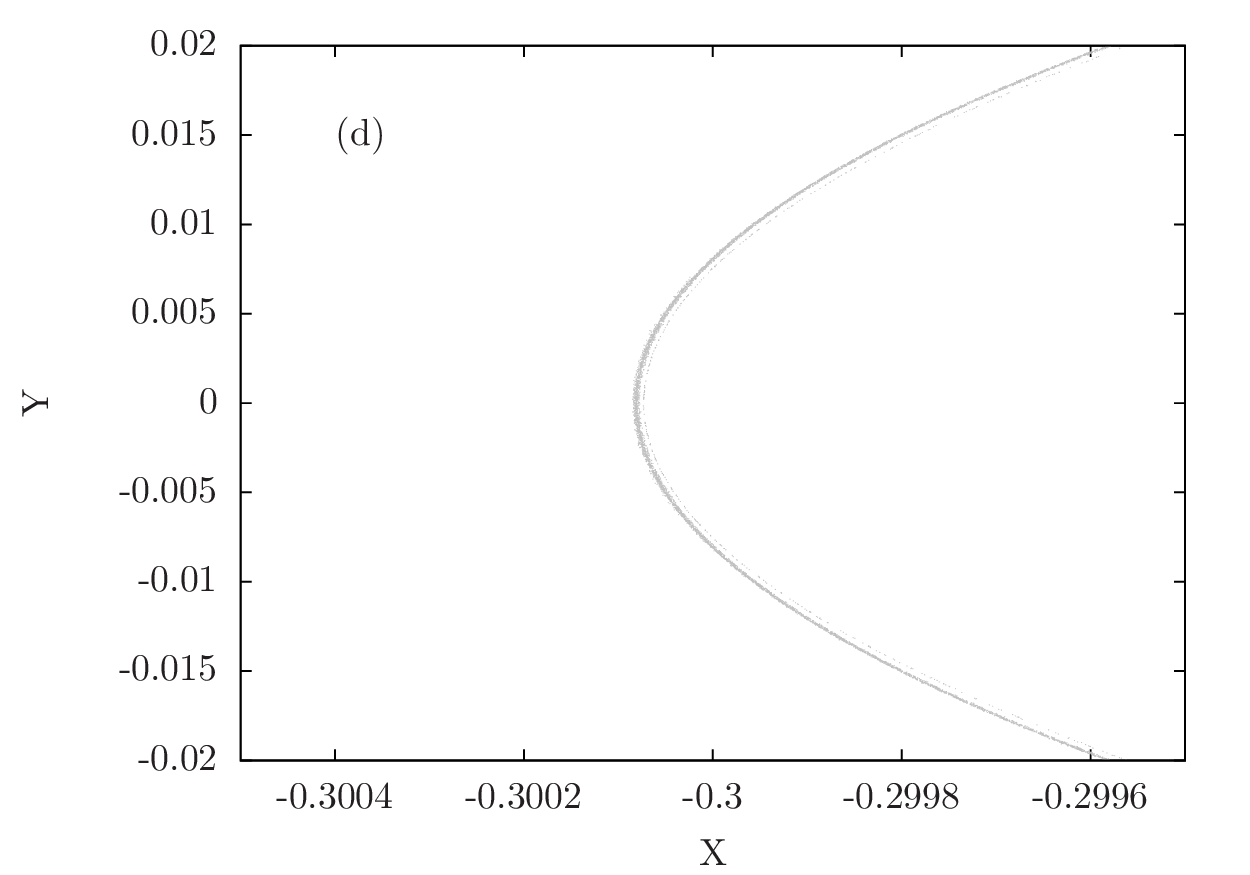}}%
  \caption{Details of a region of the $n=0$ ring for
    (a)~$\varepsilon=0.00165$, (b)~$\varepsilon=0.00167$,
    (c)~$\varepsilon=0.00168$ and (d)~$\varepsilon=0.001683$. In~(b)
    and~(c) we have distinguished bunches of particles which move as
    such, i.e., without spreading throughout the corresponding
    strand. These bunches of particles mark the occurrence of arcs or
    clumps in the ring.}
  \label{fig:ringeps}
\end{figure}

We turn now to the case of non-zero eccentricity. The explicit time
dependence of the elliptic Kepler motion cannot be removed, not even
by canonically transforming to a pulsating--rotating frame (which
defines the generalized Jacobi integral). Therefore, the Hamiltonian
system has two-and-half degrees of freedom and no constants of
motion. In this case the invariant tori do not separate regions in
phase space and new phenomena like Arnold diffusion may
appear~\citep{Arnold1964}. In spite of this, regions of trapped motion
do exist at least in the sense of effective or practical stability,
i.e., for very long but maybe not infinite
times~\citep[see][]{Jorba1997a,Jorba1997b}. We checked this by
considering a large number of consecutive bounces in the numerical
experiments (ring particles must display at least 100000 collisions
with the disk to be considered trapped), and by confirming the
appearance of a non-statistical peak in the decay-rate statistics
(see~\citealt{Merlo2004,Benet2005} for details). Therefore, the theory
can indeed be extended to systems with more degrees of freedom in
terms of (a)~the stable tori as the organizing centers of the dynamics
and, (b)~the regions of bounded motion that are found close to them
which are defined now at least on an effective stability
sense~\citep{Merlo2007}. Rings are then obtained by the construction
described above.

For extremely small values of $\varepsilon$ the ring is similar to its
counterpart for the circular case, except that it is narrower, i.e.,
the regions of trapped motion become smaller in phase space by
increasing the eccentricity. But breaking the circular symmetry has
consequences in the structural properties of the ring beyond its
narrowness. In Figs.~\ref{fig:ringeps} we plot a detail of the $n=0$
ring for slightly different values of $\varepsilon$: The rings display
multiple components (strands) and incomplete rings (arcs). Indeed,
Fig.~\ref{fig:ringeps}(a) shows a ring with two strands, the ring in
Fig.~\ref{fig:ringeps}(b) has three, and the rings of
Figs.~\ref{fig:ringeps}(c) and~(d) have again two components. The
different strands actually are entangled among each other (see
\citealt{Merlo2007}) in what reminds us the braids observed in the F
ring of Saturn~\citep{Esposito2006}. In addition, in
Figs.~\ref{fig:ringeps}(b) and (c) we observe the appearance of arcs
within the ring, which we have distinguished in the figures as the
dark grey clumps. In Fig.~\ref{fig:ringeps}(b) the arcs appear
immersed in the innermost strand, while in Fig.~\ref{fig:ringeps}(c)
that strand has disappeared allowing us to unambiguously identify the
corresponding arcs. These structures are evocative of Adam's arcs
observed in Neptune and of the sets of clumps observed in some
ringlets in Saturn~\citep{Esposito2006}. In Fig.~\ref{fig:ringeps}(c)
we have also succeed to identify another set of arcs which appear
embedded in the outermost strand; in Fig.~\ref{fig:ringeps}(d) there
are no arcs embedded in the rings.

The small difference in the eccentricity in the last three cases
indicates that the structural properties of the ring depend very
sensitively of the precise value of $\varepsilon$ in this
model. Furthermore, breaking the circular symmetry induces a strong
azimuthal dependence in the regions of trapped motion in phase space,
as manifested by the arcs. These structural properties of the ring are
related to specific phase-space aspects of systems with more than two
degrees of freedom.

\section{Phase space volume of the regions of trapped motion}
\label{sec:ps_volume}

\subsection{Two degrees of freedom and stability resonances}%

The Hamiltonian system which gives rise to rings revealing multiple
components and arcs (Figs.~\ref{fig:ringeps}) corresponds to an
explicitly time-dependent Hamiltonian, a system with two-and-half
degrees of freedom. Our aim is to characterize the regions of trapped
motion and the phase space properties where the ring particles
actually move, and as we have seen, yield rings with azimuthal
structure.

To this end, we consider a relative measure of the phase--space volume
occupied by the regions of trapped motion. For scattering systems in
particular, the volume of the regions of trapped motion depends on the
degree of development of the invariant
horseshoe~\citep{Rueckerl1994}. Therefore, the volume of the regions
of trapped motion is a function of any parameter which tunes the
development of the horseshoe or that manifests it. For the rotating
disk in a circular orbit a good choice is obviously the Jacobi
integral. Yet, $J$ is not conserved when $\varepsilon$ differs from
zero, thus becoming useless. A convenient quantity in this case is the
average time between consecutive collisions with the disk,
$\langle\Delta t \rangle$. Notice that $\langle\Delta t \rangle$ is
precisely the average return time to the surface of section. Defining
the control parameter in this way is convenient since it may be
generalized to any Hamiltonian system independently of the number of
degrees of freedom.

\begin{figure}
  \centering 
  \includegraphics[angle=270,width=8cm]{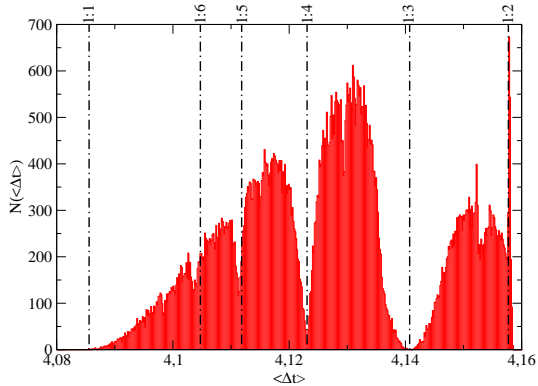}%
  \caption{Histogram of the average time between consecutive
    collisions $\langle\Delta t\rangle$ for an ensemble of particles
    of the ring of Fig.~\ref{fig:ring0}, for the rotating billiard on
    a circular orbit. The histogram represents a relative measure of
    the phase-space volume of the regions of trapped motion. The main
    gaps observed are related to the stability resonances indicated by
    dash--dotted lines. The structure displayed is universal for Hamiltonian
    systems of two degrees of freedom.}
  \label{fig:psv0}
\end{figure}

We use $t_i^{(k)}$ to denote the time of the $i$-th collision with the
disk for a particle labeled by $k$ that remains trapped, and
$\langle\Delta t^{(k)} \rangle = ( t_N^{(k)} - t_0^{(k)})/N$ its
average collision time after $N$ bounces; we further average over many
initial conditions of the ring particles ($k\gg 1$) to obtain
$\langle\Delta t \rangle$. Numerically, we consider that the initial
conditions of a ring particle belong to a region of trapped motion if
the particle displays more than $10000$ collisions with the disk; we
used the next $N\ge 17500$ collisions for the
statistics. Figure~\ref{fig:psv0} shows the frequency histogram of
$\langle\Delta t \rangle$; the corresponding region of trapped motion
yields the ring of Fig.~\ref{fig:ring0}.

The structure shown in Fig.~\ref{fig:psv0} is universal for
Hamiltonian systems with two degrees of
freedom~\citep{Contopoulos2005,Dvorak2005}. Universality implies that
it holds for a wide class of systems. The characteristic feature of
Fig.~\ref{fig:psv0} is the occurrence of specific values of the
parameter where an abrupt reduction of the phase-space volume of the
regions of trapped motion is observed, henceforth referred as
gaps. Gaps are observed at every scale and display self-similar
structure~\citep{Simo2006,Simo2007,SimoVieiro2007}. The abrupt changes
in phase space can be understood, for two degrees of freedom, from the
existence of an outer invariant curve which serves as barrier, thus
confining the motion of a region in phase space. Small changes of the
parameter destroy such invariant curves, which thus allow the
particles within certain region in phase space to escape. This
explains the sudden variation in the phase-space volume of the regions
of trapped motion observed. See \citet{Simo2007} for a careful
analysis of these aspects in H\'enon's map.

Figure~\ref{fig:psv0} provides a {\it global} description of the
region of trapped motion in phase space for $n=0$, i.e., it includes a
continuous interval of $J$ and regions of phase space that do not need
to be close to each other. An accurate estimation for the location of
the gaps is provided by the stability resonances. The stability
resonances are defined by the occurrence of a resonant condition (a
rational ratio of $2\pi$) of the linear stability exponents of the
central linearly stable periodic orbit. Therefore, they are related to
the {\it local} stability properties of the central stable periodic
orbit~\citep{Benet2008}. Note that the stability resonances must not
correspond to mean--motion resonances, which are defined as a rational
ratio between the period of the central stable periodic orbit and the
period of the disk.

For the rotating disk on a circular orbit the stability of the radial
collision periodic orbits is obtained using Eq.~(\ref{eq:stab}). We
write the eigenvalues of the linearized dynamics $D{\cal P}_J$ for the
(linearly) stable radial collision periodic orbit as $\lambda_\pm =
\exp[\pm i \eta_\pm]$. The stability resonances are thus defined by
the condition $\eta / (2\pi)= p/q$ with $p$ and $q$ incommensurate
integers. From the definition of the (imaginary) eigenvalues $\lambda$
it follows $\cos\eta={\rm Tr\,} D{\cal P}_J/2$, which can be written
in terms of the collision time $t_{\rm col} = \Delta\phi/\omega_d$
using Eq.~(\ref{eq:stab}). Notice that the eigenvalues $\lambda_\pm$
are related to each other by complex conjugation; by consequence, the
$p:q$ resonance is related to the $q-p:q$ resonance. In
Fig.~\ref{fig:psv0} we have indicated the location of some lower-order
stability resonances as dash--dotted vertical lines. The
correspondence with the gaps is excellent, though not perfect, as
illustrated by the 1:5 or 1:6 stability resonances.
The reason that there is no perfect correspondence of the gaps and the
stability resonances is related to the fact that the latter are
defined using a local estimate of the stability of the central
(linearly) stable periodic orbit. Changing slightly the control
parameter changes accordingly the stability properties of this
periodic orbit. In turn, the phase-space volume of the regions of
bounded motion is determined, as mentioned earlier, by the outer
invariant curve which bounds the region. The existence of this
invariant curve depends subtly on the specific value of the control
parameter and thus a change in the control parameter has a global
effect.

\begin{figure}
  \centering
  \includegraphics[angle=0,width=12cm]{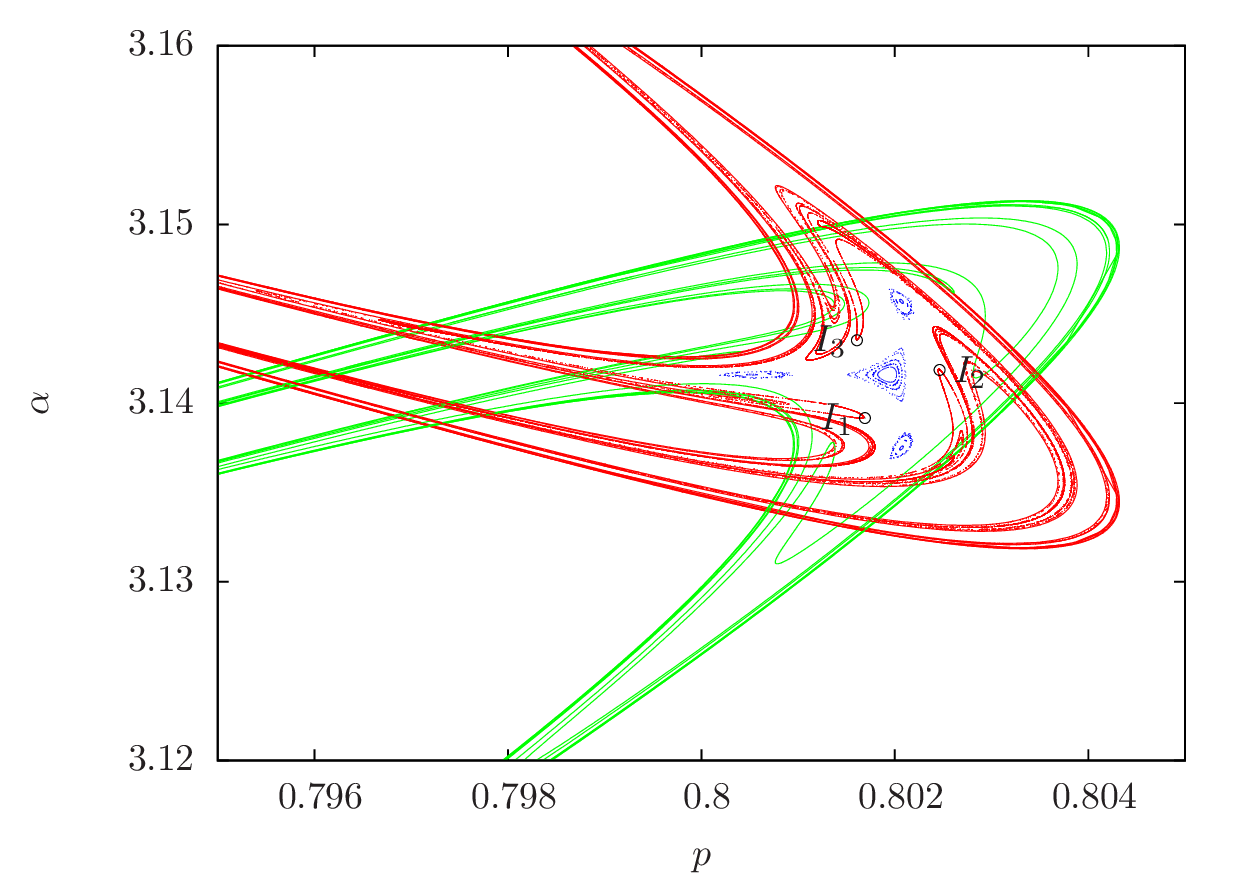}
  \caption{ Surface of section for the billiard system on a circular
    orbit with $J/(R-d)^2=0.293011$. The labels indicate subsequent
    tips of the unstable manifold. Note that these tips mimic the
    periodicity of the period-three unstable periodic orbit associated
    with the secondary islands displayed.}%
  \label{fig:manifolds}%
\end{figure}

It is interesting to note that using a local estimate we obtain a good
approximation for the location of the gaps, at least in the case of
low order stability resonances, recalling that these gaps uncover
global aspects of the phase space. To understand how local estimates
can provide knowledge on global aspects of the dynamics, we observe
that the stability resonances are related to rational winding numbers
close to the (linearly) stable periodic orbit. In turn, the
phase-space volume of the regions of trapped motion are ultimately
related to the invariant manifolds of the unstable periodic orbit and
their homoclinic connections, i.e., the underlying horseshoe
structure. The connection among these aspects, for low developed
horseshoes, can be established by the relation between the {\it formal
  development parameter} which characterizes the horseshoe
development~\citep{Jung1999} and the characteristic period of an
outermost stable periodic orbits~\citep{Jung2004}. In particular,
notice that certain segments of the manifolds of the unstable periodic
orbit (at a locally turning point of the manifolds or tips) are quite
close to the unstable periodic orbits that are the companions of the
outermost secondary islands, thus mimicking the periodicity of such
islands. For low order stability resonances where the change in the
phase space volume occupied by trapped trajectories is very drastic,
these manifolds do enter deep enough the horseshoe structure and
become very close to the linearly stable periodic orbit. This
describes the connection of a global property of the phase space with
local estimates, which is illustrated in Fig.~\ref{fig:manifolds}
close to the minimum of the $1:3$ stability resonance. Conversely, if
the invariant manifolds lie somewhat outside the surroundings of the
stable periodic orbit, the estimate becomes less accurate.

From Fig.~\ref{fig:manifolds}, we obtain that the formal development
parameter is $\beta=1/4$. This value of $\beta$ is related with the
period $T_\beta=3/2 - \log_2\beta=3.5$ given in units of the average
return time to the surface of section (see~\citealt{Jung2004}). As
noted in \citealt{Jung2004}, $T_\beta$ is an average period with an
associated error of $\pm 0.5$. As we approach further the minimum of
the histogram associated with the $1:3$ stability island, $\beta$
increases slowly and $T_\beta$ approaches the expected value of three,
which corresponds to $\beta=2^{-3/2}$. Notice that this value is well
beyond the value $\beta=3/8$, which is known to have problems with
respect to a consistent definition of the development
parameter~\cite{Rueckerl1994}. A more detailed study of this will be
presented elsewhere.

As mentioned above, for systems with two degrees of freedom the
structure of the phase-space volume occupied by regions of trapped
motion (Fig.~\ref{fig:psv0}) is universal and displays self-similar
structure. Universality is a consequence of generic aspects of the
dynamics and of the normal form~\citep{Arnold1989,Gelfreich2002}. In
particular, the passage through the 1:3 stability resonance leads
universally to local instabilities which decrease violently the
phase-space volume occupied by trapped orbits, yielding a very small
but not necessarily zero phase-space volume of the region of trapped
motion. Resonances of order higher than 4 do not induce such strong
local instabilities; for the 1:4 stability resonance the behavior
depends upon which contribution, resonant or non-resonant, dominates
the normal form~\citep[see][]{Arnold1989}. These results follow from
the structure of the resonant normal form, i.e., the relevant
nonlinear contributions to the stability analysis of a central
linearly stable periodic orbit, and the scenario of development of the
horseshoe.

\subsection{Beyond two degrees of freedom: Strands and Arcs}

\begin{figure}
  \centering
  \includegraphics[angle=270,width=8cm]{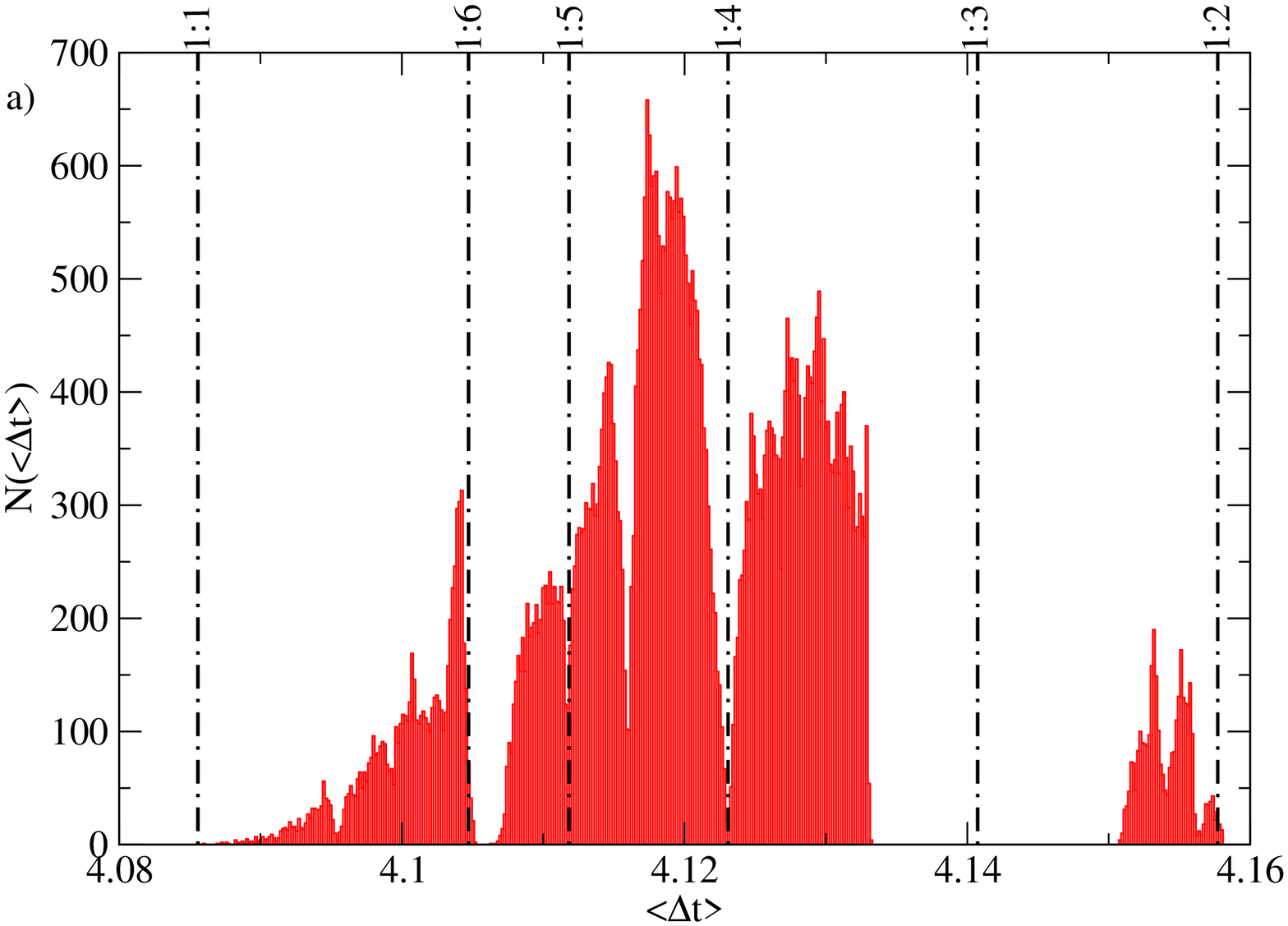}
  \includegraphics[angle=0,width=8cm]{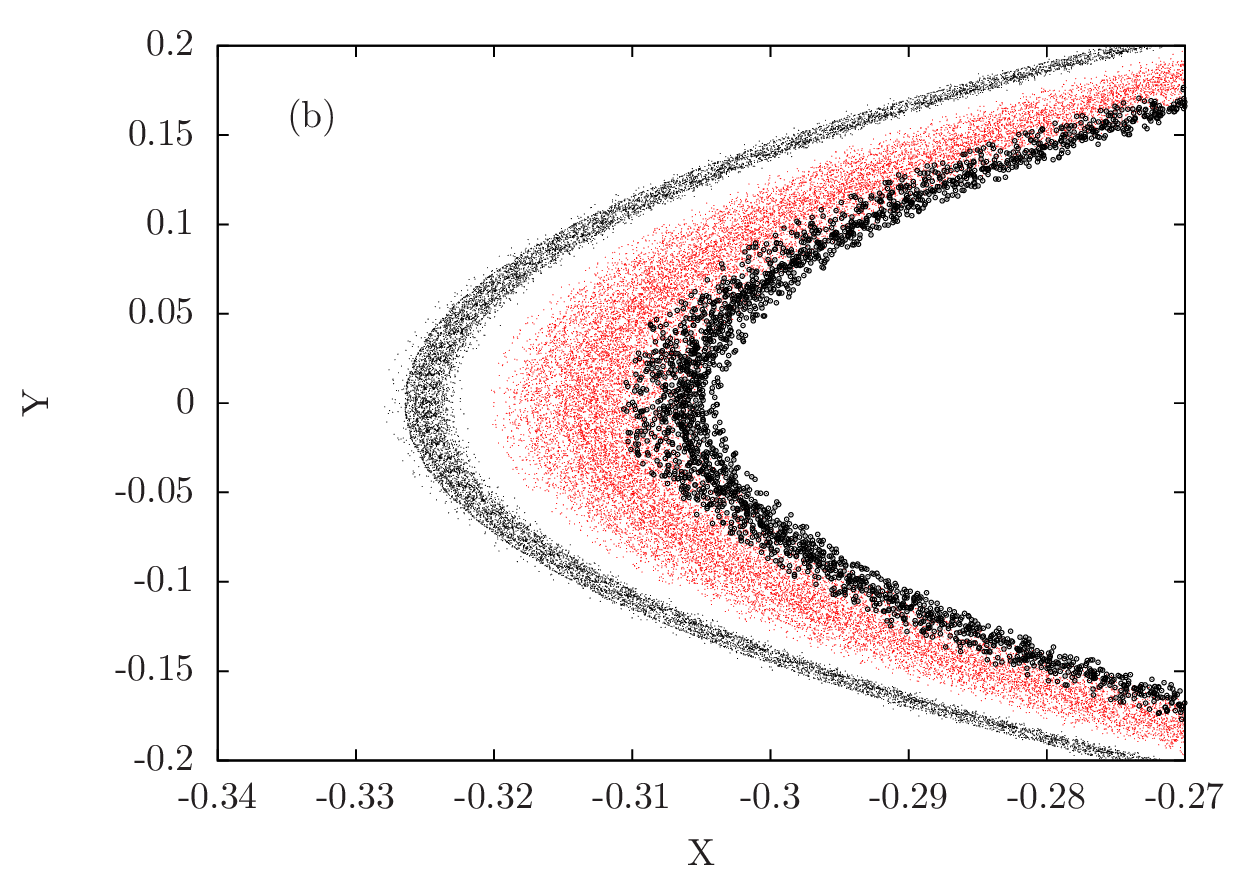}%
  \caption{ (a)~Same as Fig.~\ref{fig:psv0} for the billiard on an
    elliptic Kepler orbit for $\varepsilon=0.0001$. The resonances
    indicated correspond to $\varepsilon=0$. (b)~Detail of the
    corresponding ring. The gap of the 1:3 stability resonance is
    wider, which causes a division of the ring thus forming a
    two-component ring. The innermost component of the ring
    corresponds to the region on the left of the 1:6 stability
    resonance, and the middle one to the region between the 1:6 and
    1:3 resonances.}%
  \label{fig:psv1}%
\end{figure}

The gaps defined by the stability resonances at first sight seem not
to be important for the structure of the ring. For the disk on a
circular orbit, they do manifest in the ring's density profile. For
non--zero $\varepsilon$ the relative measure of the phase-space volume
occupied by regions of trapped motion displays new features, and this
change is manifested in the structure of the corresponding
ring. Figure~\ref{fig:psv1}(a) illustrates the structure of the
phase-space volume of the regions of trapped motion for
$\varepsilon=0.0001$. Note that the gaps corresponding to the 1:3 and
1:6 stability resonance are wider and divide the histogram in three
regions. We assign different plotting characters to initial conditions
belonging to these regions; Fig.~\ref{fig:psv1}(b) illustrates the
ring obtained. The ring displays two components, separated by a ring
division, which are entangled and form a braided ring.

\begin{figure}
  \centerline{
    \includegraphics[angle=270,width=7cm]{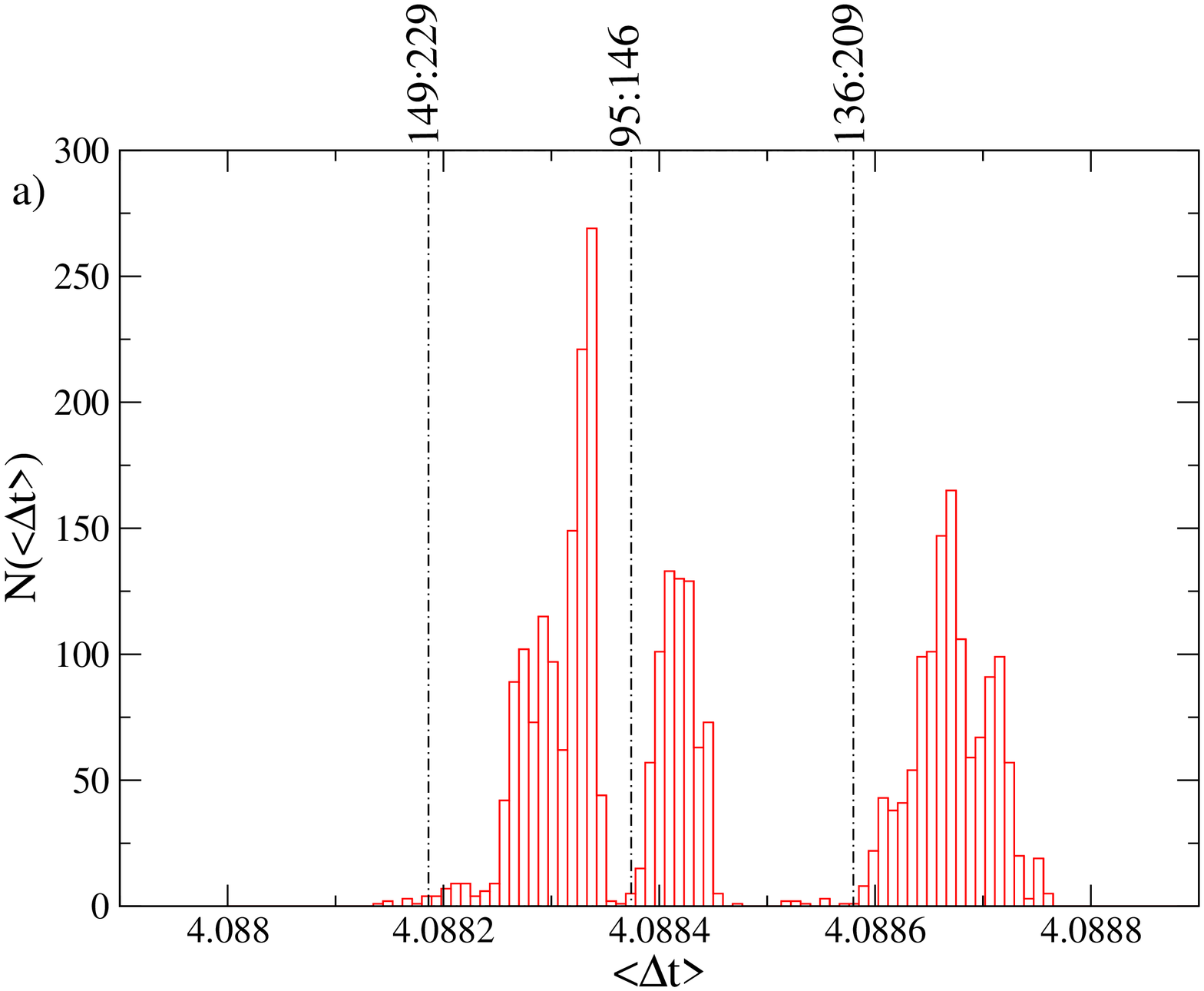}
    \includegraphics[angle=270,width=7cm]{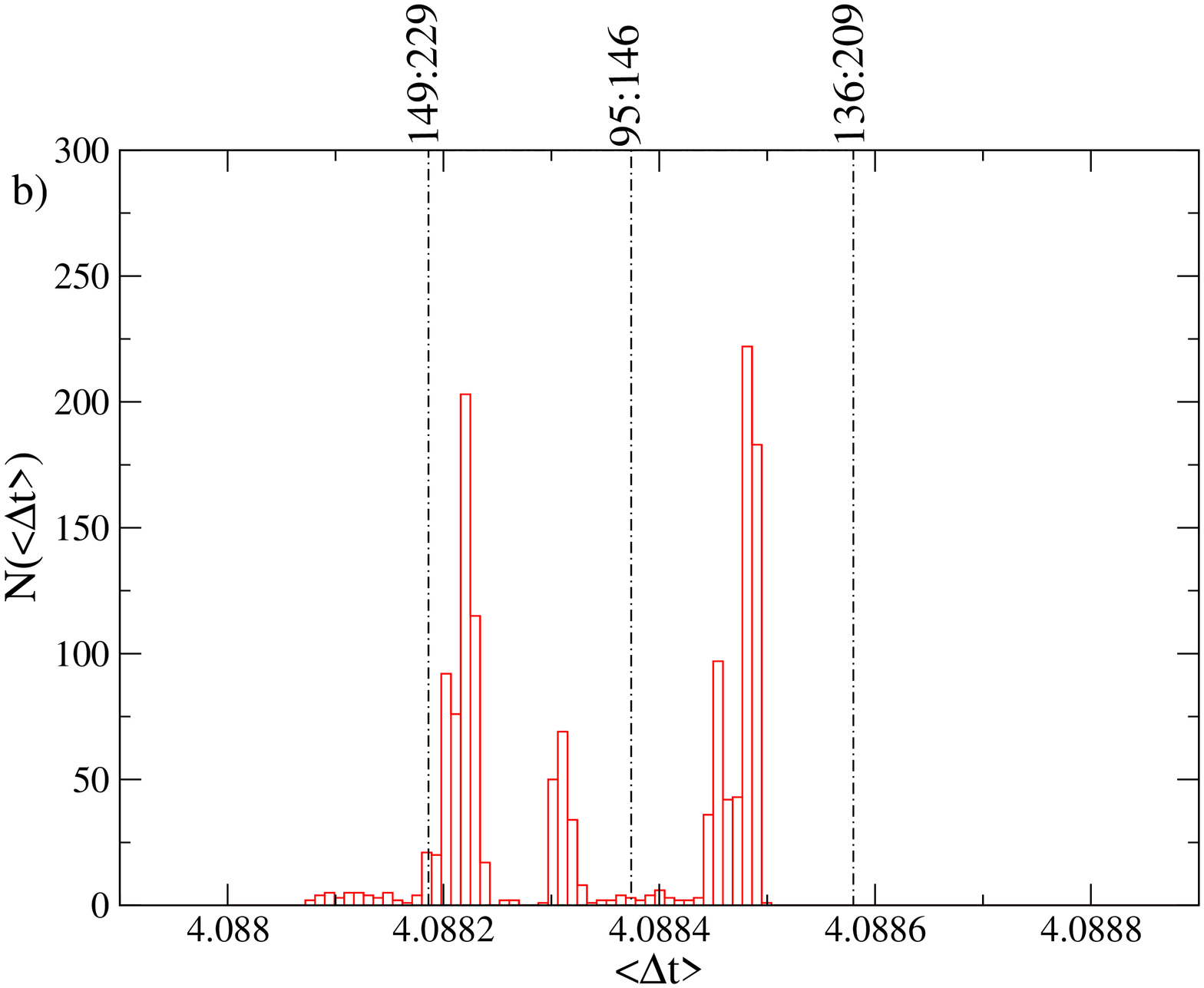}}%
  \centerline{
    \includegraphics[angle=270,width=7cm]{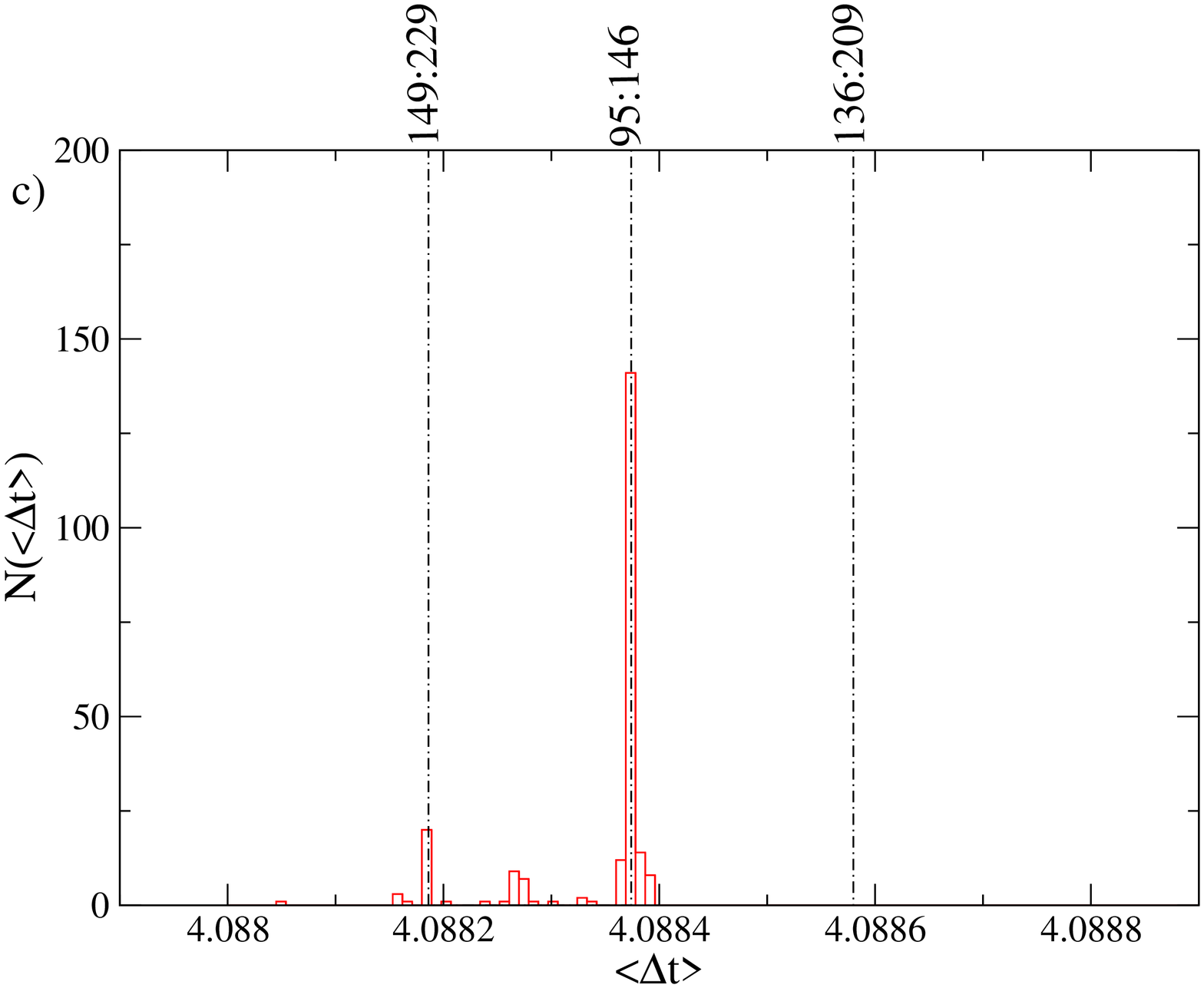}
    \includegraphics[angle=270,width=7cm]{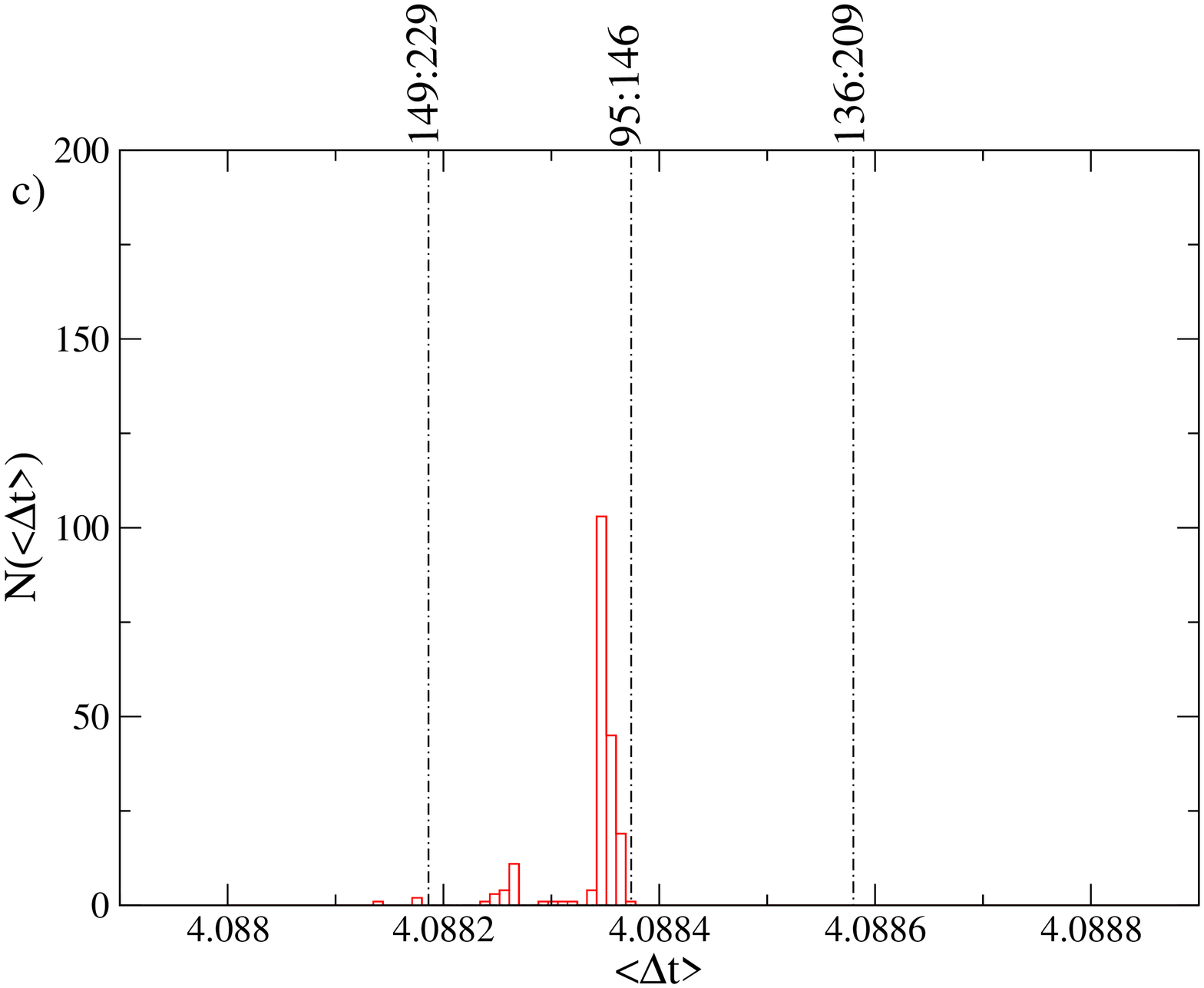}}%
  \caption{Same as Fig.~\ref{fig:psv0} for (a)~$\varepsilon=0.00165$,
    (b)~$\varepsilon=0.00167$, (c)~$\varepsilon=0.00168$ and
    (d)~$\varepsilon=0.001683$, which correspond to the rings given in
    Fig.~\ref{fig:ringeps}. Some mean-motion resonances occurring in the
    interval of $\langle\Delta t\rangle$ are indicated by dashed--dotted 
    lines. Some of these are related to the appearance of arcs.}
  \label{fig:psveps}
\end{figure}

We observe that each ring component is related to a different region
of the histogram. This is a consequence of the large gap opened by the
1:3 stability resonance, which effectively separates the phase space
regions of trapped motion in two disjoint
regions~\citep{Benet2008}. Yet, while the 1:6 stability resonance
widens up its corresponding gap, it does not separate {\it enough} the
phase-space regions around it. The rings formed separately (by
projection into the $X-Y$) of these regions manifest a clear
segregation of ring particles, except for a thin strip. In
Figs.~\ref{fig:psveps} we present the relative measures of the phase
space volume occupied by the regions of trapped motion for the values
of $\varepsilon$ corresponding to the rings shown in
Fig.~\ref{fig:ringeps}. As in the case $\varepsilon=0.0001$, the
stability resonances induce instabilities which widen up the gaps of
the case $\varepsilon=0$. This destroys locally the trapping mechanism
in certain regions in phase space and yields one or more divisions of
the ring. Therefore, multiple ring components are obtained by exciting
the stability resonances through non-zero eccentricity. Note that
these structural properties follow from the higher dimensionality of
phase space and nonlinear effects, i.e. $\varepsilon\neq 0$.

The histograms of $\langle\Delta t\rangle$ representing a relative
measure of the phase-space volume of the regions of trapped motion and
the stability resonances clarify the origin of the multiple components
of the ring but do not explain the occurrence of arcs. To understand
their origin, we first observe that there are {\it exactly} 149 arcs
along the whole ring for $\varepsilon=0.000167$,
(Fig.~\ref{fig:ringeps}(b)). Secondly, the exact angular configuration
of the arcs (numbered in an arbitrary way with respect to one ring
particle) is repeated after 229 bounces with the disk. This
observations leads us to suspect that the appearance of arcs is
related to the occurrence of mean-motion resonances. To check this, we
have plotted in Figs.~\ref{fig:psveps} the position of the 149:229
mean-motion resonance, i.e., $\langle\Delta t\rangle /(2\pi) =
149/229$. We note that for $\varepsilon=0.000168$ the histogram of the
phase space volume of the region of trapped motion indicates that
there is a region of trapped motion, which is quite narrow, precisely
agreeing with the condition for the mean-motion resonance. In this
case, arcs are clearly manifested in Fig.~\ref{fig:ringeps}(c). The
same analysis indicates that these arcs are expected for
$\varepsilon=0.000167$ since this mean-motion resonance is embedded
within a region of trapped motion, see Fig.~\ref{fig:psveps}(b). We
indeed confirmed this expectation as illustrated in
Fig.~\ref{fig:ringeps}(b). The fact that the region of trapped motion
is somewhat wide (in the $\langle\Delta t\rangle$ interval) is the
reason for the ring component to be covering the arcs. A similar
analysis can be carried out for other resonances. In particular, the
95:146 resonance also occurs inside the $\langle\Delta t\rangle$ range
of interest. We find that this resonance shows up for
$\varepsilon=0.000168$, but is embedded within a region of trapped
motion, see Fig.~\ref{fig:psveps}(c). As shown in
Fig.~\ref{fig:ringeps}(c), in this case the arcs are also embedded
inside a ring component.

\begin{figure}
  \centering 
  \includegraphics[angle=270,width=8cm]{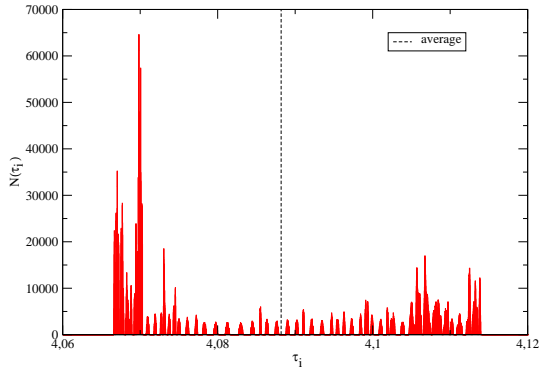}%
  \caption{Histogram of the time between two consecutive collisions
    $\tau_i$ with the disc for one initial condition which belongs to
    an arc. The average time is indicated by a dashed line.  }
  \label{fig:histarcs}
\end{figure}

In Fig.~\ref{fig:histarcs} we plot the histogram of elapsed time
between consecutive collisions, $\tau_i=t_i-t_{i-1}$, for one initial
condition that belongs to a specific arc. The figure shows a number of
bumps which are clearly separated from each other, and their average
value. The histogram demonstrates that the motion of a ring particle
within an arc is more constrained than within a ring component, which
is consistent with the restriction imposed by the mean-motion
resonance condition. 

To have a physical picture of the phase space structure we define the
Poincar\'e map from the dynamics on a surface of section which
corresponds to collisions with the disc. A periodic orbit in a $p:q$
mean-motion resonance is a fixed point of $q$ iterations of the
Poincar\'e map. In the case of circular orbit of the disk, the
symmetric case, a whole family of these periodic orbits exist and can
be parameterized by the angle $\phi$ where the particle hits the disc.
For small non-vanishing $\varepsilon$ the $p:q$ family of periodic
orbits persist. Depending on the location in the orbit of the disk
$\phi(t)$ where the bounce takes place the orbits may be stable or
unstable~\citep{Merlo2004,Benet2005}, i.e., there is an azimuthal
dependence in the stability properties of the periodic
orbits. Unstable periodic orbits yield either a thin layer of chaotic
motion within the region of trapped motion, or escaping channels. The
stable ones define individual secondary regions of bounded motion,
which are separated by the unstable orbits and their associated
invariant structures, the projection of each one yielding an
individual arc. Figure~\ref{fig:histarcs} is a different view of this
effect. These secondary regions of bounded motion may be embedded
within the main one or may appear as satellite islands of
stability. In the former case the arcs overlap with the main strands
(cf. Figs.~\ref{fig:ringeps}(c) and \ref{fig:psveps}(c)); in contrast,
in the latter case some arcs may show up as completely separated from
strands (cf. Figs.~\ref{fig:ringeps}(c) and \ref{fig:psveps}(c)). The
scenario described above resembles the typical Poincar\'e-Birkhoff
type of structure found in Hamiltonian systems of two degrees of
freedom, at least on an effective stability sense for more than two
degrees of freedom. We emphasize that arcs manifest an azimuthal
dependence in the structure of the ring, which can be understood in
these terms.

The above considerations lead us to conclude that two condition must
be satisfied for the occurrence of arcs: First, a mean-motion
resonance condition must be met, which in a way allows for the
appearance of Poincar\'e-Birkhoff-like phase-space structures and,
secondly, the regions of trapped motion should be thin enough. The
last condition is needed to distinguish the arcs from the main ring
components. In principle, any high-order mean motion resonance may be
observable as a chain of arcs. Yet, numerically, we deal with a finite
number of particles distributed in a rather extended region of a
higher dimensional phase space. Then, it is a difficult task to
distinguish from the structure of the obtained ring if the particles
move in individual packets that form the arcs or within a connected
ring component.

\begin{figure}
  \centering 
  \includegraphics[angle=0,width=12cm]{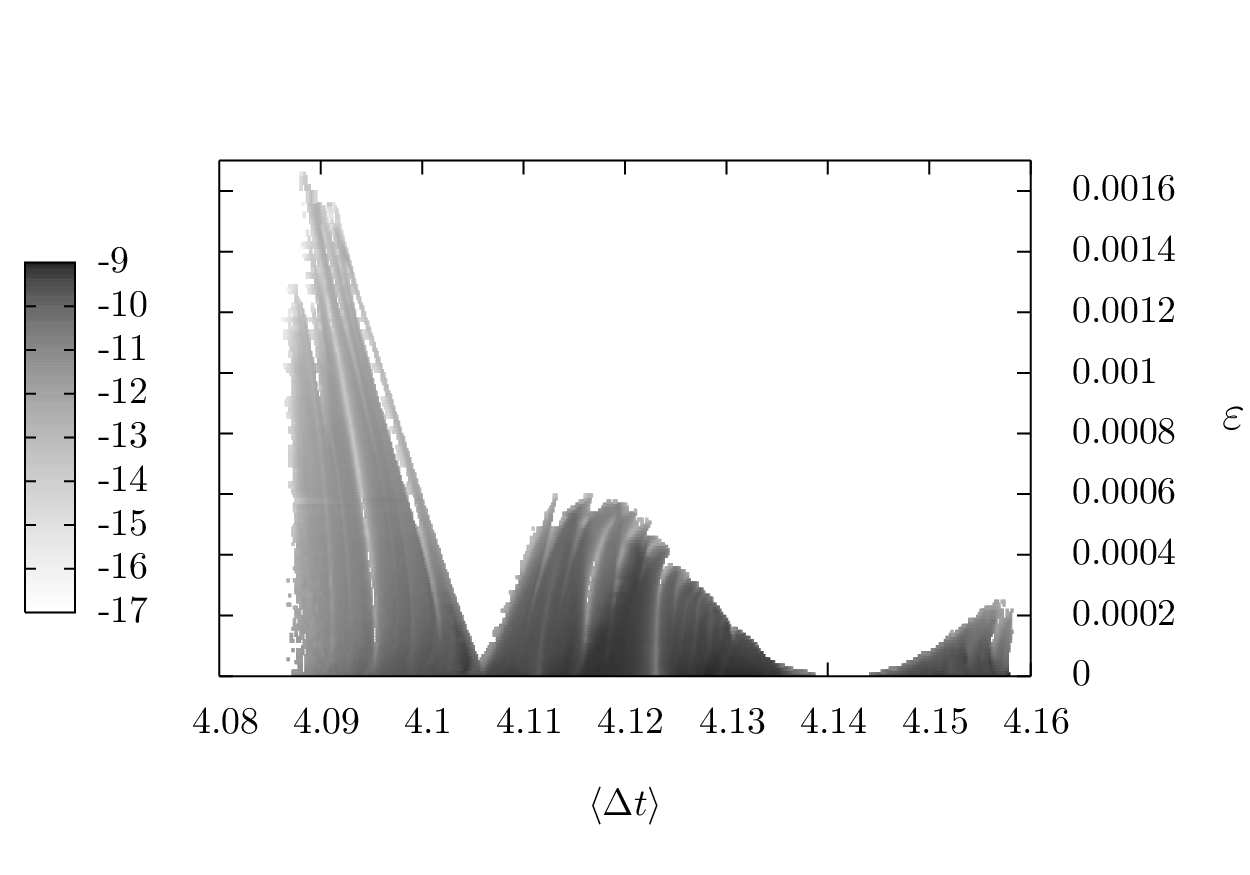}%
  \caption{Contour plot for the representation of the logarithm of the
    volume of phase space occupied by regions of trapped motion, in
    terms of the eccentricity $\varepsilon$ and $\langle\Delta
    t\rangle$. For fixed $\varepsilon$, the histogram was normalized
    to the actual value of the volume of phase space occupied
    by regions of trapped motion.}%
  \label{fig:psvT}
\end{figure}

We finish this section showing the combined dependence upon
$\varepsilon$ and $\langle\Delta t\rangle$ of the phase-space volume
of the regions of trapped motion. Figure~\ref{fig:psvT} displays this
dependence as a contour plot of the logarithm of the phase-space volume
of the regions of trapped motion. For each $\varepsilon$ computed, we
normalized the corresponding histogram to the actual value of the
phase-space volume occupied by regions of trapped motion for that
case. Typically we have used about 50000 initial conditions to obtain
the histogram for a given $\varepsilon$, except for higher
eccentricities where we have used a smaller number due to the immense
time required to obtain a representation of the higher-dimensional
phase-space region in question.

Figure~\ref{fig:psvT} manifests the evolution of individual strands
when the eccentricity is changed. For a specific strand, the grey
tones represent the relative phase-space volume as a function of the
eccentricity. The plot also shows which of the stability resonances of
the circular case are excited, and actually the whole evolution of
these resonances in terms of $\varepsilon$. Clearly
Fig.~\ref{fig:psvT} provides a global picture in parameter space of
the phase-space volume of the regions of trapped motion. Using it
allows, among other, to search for $\varepsilon$ values where arcs
manifest.

\section{Summary and conclusions}
\label{sec:concl}

In this paper we have studied the phase-space volume of the regions of
trapped motion in a specific Hamiltonian system of two and
two-and-half degrees of freedom depending on a parameter, an
impenetrable hard disk whose center moves on a Kepler orbit with fixed
eccentricity~\citep{Benet2004,Merlo2007}. We have also related the
phase-space volume of the regions of trapped motion to the structural
properties of the patterns obtained by projecting into the $X-Y$
plane, at a fixed time, the phase-space locations of an ensemble of
particles whose initial conditions belong to these regions of trapped
motion. These patterns form ring structures similar to the observed in
narrow planetary rings, being non-circular and displaying
sharp-edges. These structural properties follow directly from the
scattering approach~\citep{Merlo2007}.

For non-zero eccentricity, the rings may display strands and
arcs. Multiple ring components are due to the separation of
phase-space regions of trapped motion, obtained by widening up the
gaps observed in the histograms representing the phase-space volume of
the regions of bounded motion. The location of these gaps is
approximated by a resonant condition on the stability exponents of the
linearized dynamics around a central linearly-stable periodic
orbit. This local estimate yields very good results for the low order
resonances, while for higher ones the approximation is less accurate; this
is due to global effects related to outer invariant curves that define
the region of bounded motion, which are not well described by the
local estimate. Arcs are observed whenever an additional resonant
condition is imposed: a mean-motion resonance. This result represents
an independent confirmation that such resonant mechanisms indeed yield
corotation sites in phase space~\citep{Goldreich1986}. This is in
agreement with the current understanding of Adam's arcs in
Neptune~\citep{Porco1991,Namouni2002}, which require to take into
account the gravity of the arcs to match recent
observations~\citep{Sicardy1999,Dumas1999}. Actually, our results show
that different specific resonances can coexist, and thus yield arcs
with different angular widths. We emphasize that both, strands and
arcs, are obtained from setting the initial conditions for the ring
particles in an interval that contains the regions of trapped motion,
and are not imposed artificially. These structural properties are
consequences of the high-dimensional and non-linear aspects of the
dynamics, in the sense that individual strands and arcs do not occur
for Hamiltonian systems with two degrees of freedom. The fact that we
obtain them in such a non-realistic system like the hard-disk moving
on a Kepler orbit make us confident of the applicability of these
ideas in realistic situations.

Our results provide a complete representation of the phase-space
volume of the regions of trapped motion for a Hamiltonian system of
more than two degrees of freedom. The natural question that arises is
whether its structure and properties hold universally for systems with
more than two degrees of freedom. In particular, we may ask if the
gaps associated with the 1:3 and the 1:6 stability resonances are {\it
  always} excited first with respect to variations of the relevant
symmetry-breaking parameter, and what intrinsically distinguishes the
instability of the 1:6 gap with respect to the 1:5, or put
differently, why do other gaps not separate so efficiently. Another
interesting question is related to improve the simple estimates given
by the stability resonances for the location of the gaps in the
histograms of the phase-space volume occupied by trapped orbits,
especially for the case of higher-order stability resonances. These
questions remain open. We do think that the behavior of the 1:3
stability resonance holds in general: The 1:3 stability resonances are
optimally excited by the symmetry-breaking parameter, and we expect to
have a sensitive separation of the regions of trapped motion in phase
space under small changes of the symmetry--breaking parameter. This is
based on the particularly deep gap observed around this resonance for
the case where the symmetry holds (two degrees of freedom).

The fact that the rings of the scattering billiard on a Kepler orbit
display qualitative similarities to the narrow planetary rings
consistently is encouraging: The scattering approach, which is based
on the relevant phase-space structures, yields structured rings in
qualitative agreement with the observations. The motivation to
continue along these lines for systems with more than two degrees of
freedom, for example, is to build a theory upon which we can determine
or predict the limits of stability that define the regions of trapped
motion, and then obtain semi-analytical estimates for the edges of the
ring.

\begin{acknowledgements}
  We acknowledge financial support provided by the projects IN--111607
  (DGAPA--UNAM) and 79988 (CONACyT). We are thankful to \`Angel Jorba
  and Carles Sim\'o for discussions and comments, and Fathi Namouni for
  critical remarks. O.~Merlo is a postdoctoral fellow of the Swiss
  National Foundation (PBBS2--108932).
\end{acknowledgements}


\end{document}